# Computational complexity of vacua and near-vacua in field and string theory

James Halverson
*Department of Physics, Northeastern University, Boston, Massachusetts 02115-5000 USA*

Fabian Ruehle
*Rudolf Peierls Centre for Theoretical Physics, University of Oxford,
Parks Road, Oxford OX1 3PU, United Kingdom*

We demonstrate that the problems of finding stable or metastable vacua in a low energy effective field theory requires solving nested nondeterministically polynomial (NP)-hard and co-NP-hard problems, while the problem of finding near-vacua can be solved in polynomial (P) time. Multiple problems relevant for computing effective potential contributions from string theory are shown to be instances of NP-hard problems. If $P \neq NP$, the hardness of finding string vacua is exponential in the number of scalar fields. Cosmological implications, including for rolling solutions, are discussed in light of a recently proposed measure.

## I. INTRODUCTION

The enormous landscape of string theory realizes rich and diverse physical phenomena, but poses critical practical obstacles to its study. The number of topological choices for the string background is already gargantuan. The estimated size [1] is $10^{272,000}$ for flux vacua [2] and a lower bound of $10^{755}$ for F-theory geometries [3] (see also [4]), which necessitates the use of modern techniques from computer science, e.g., from data science and machine learning [5–9]. On top of the topological background choice, each string background typically comes with a set of moduli and other scalar fields. Their potential, which is fixed in terms of the background data, is generated from both perturbative and nonperturbative effects and contains in general a vast number of extrema.

However, size is not the only obstacle: string vacuum studies are faced with issues of both computational complexity [10] and undecidability [11]. There are (physical) minimization problems that require computation time that is exponential in the input size (unless P = NP) or cannot be decided, respectively, by an algorithm on a classical computer. We refer to problems that take exponential time as *hard*. In general, complexity and decidability have dramatic consequences for concrete studies in the landscape. We emphasize, however, that hardness does not immediately imply intractability, since exponential time may be affordable for concrete instances of problems at parameter values relevant in string theory (see, e.g., [12]). Beyond these practical issues, complexity is also known to affect the dynamics of physical systems that realize a landscape of metastable states, for example in spin glasses and protein folding; see [10,13] for discussions. It is therefore natural to ask whether complexity also plays a role in the quantum dynamics and measure of the string landscape, as recently proposed in [14].

With this practical and potential dynamical relevance in mind, we study the computational complexity of determining vacua and near-vacua in both quantum field theory and string theory. The practical issue is primarily motivated by the general importance of finding vacua, but also by the recent conjecture [15] that the string landscape does not exhibit any de Sitter (dS) vacua; see also [16,17] for further investigations. In the latter context, the existence of dS vacua requires next-to-leading order contributions to avoid a no-go theorem of Maldacena-Nuñez [18], as emphasized in [19]. The computation and systematic control over such corrections is quite difficult (as is the subsequent minimization procedure), but is the central focus of both the KKLT [19] and LARGE volume [20] moduli stabilization scenarios; for concrete models in these scenarios that exhibit stabilized vacua under reasonably clear assumptions, see, e.g., [21]. The present work investigates whether this naive difficulty experienced in attempting to concretely construct vacua can be made more rigorous, by casting it into the language of computational complexity.

Our main result is that the computational hardness of metastable string vacua arises not only from the hardness of determining contributions to the effective scalar potential, but also from the hardness of actually finding metastable minima; i.e., one is faced with nested computationally hard problems. This explains the dearth of concrete studies of string vacua at large numbers of scalar fields (e.g., Calabi-Yau compactifications with $h^{11}$ or $h^{21}$ of $\mathcal{O}(10)$ or higher), despite the expectation that most of the landscape lies in such regions. In a recently proposed cosmological measure that utilizes computational complexity, our results imply that the universe may not have found a local minimum. For a discussion of this possibility, important and interesting caveats, and also practical implications, see the discussion.

This paper is organized as follows. In Sec. II we review computational complexity. In Sec. III we study the complexity of finding vacua and near-vacua of a scalar potential, focusing on the hardness of both the optimization and decision versions of the problems. In Sec. IV we study the complexity of determining certain important scalar potential contributions in string theory. Many are instances of known NP-complete problems. In Sec. V we discuss the practical and physical implications of our results. Appendix contains a proof relevant for the hardness of minimizing the scalar potential.

## II. COMPUTATIONAL COMPLEXITY

In this section we review basic elements of computational complexity. For further introduction, see [10] in the physics literature, [22] for an optimization account, and [23] for an in-depth account.

We define a PROBLEM $F: I \to B$ as a map from instances to outputs. For example, $I$ could be a set of $C^\infty$ functions $f: \mathbb{R} \to \mathbb{R}$ and $F(f \in I)$ could be its global minimum. Using this, a DECISION PROBLEM is a problem where $B = \{\text{yes}, \text{no}\}$. An example of a decision problem is CLIQUE. A clique of an undirected graph $G$ is a set of vertices that are all connected to one another, and we will refer to a clique with $n$ vertices as an $n$-clique. CLIQUE is the problem:

CLIQUE: Given an undirected graph $G$ and $n \in \mathbb{Z}$, does $G$ have an $n$-clique?

Here $I = S \times \mathbb{Z}$, where $S$ is the set of undirected graphs.

By an *algorithm* (on a classical computer) we mean a computational procedure that can be modeled by a deterministic Turing machine.[1] An algorithm computes $F$, and different algorithms may have different runtimes. A polynomial time, or *polytime* algorithm produces $F(x \in I)$ in an amount of time bounded by a polynomial in the input size. We will refer to any algorithm that is not polytime as exponential time, which means that its execution time is at least exponential.

We focus on some of the most-studied complexity classes. P is the class of problems with polytime solution algorithms. NP is the class of problems with polytime verifier; i.e., yes-instances can be verified in polytime for a given input (by a so-called witness or certificate). There also exist other definitions that do not rely on the existence of a witness. Perhaps the most famous problem in complexity theory is to determine whether or not $P = NP$. A strong consensus is that $P \neq NP$, but the problem is open.

It should be noted that each time we state the complexity class P, we mean that the number of operations needed to solve the problem is bounded by a polynomial function of the input size. The time it takes to perform each operation itself will grow with the precision at which we perform the computation. For example, adding two $d$-digit numbers is one operation, but takes $d$ sub-additions (each digit is added individually). If each subaddition takes a time $t_{\text{sub}}$, the time it takes to compute at $d$ digit precision will be $d \cdot t_{\text{sub}}$, which can make even problems in P unfeasible at very high precision.

There is a notion of the hardest NP problems. A problem $G$ is NP-hard if there exists a polytime reduction (henceforth, reduction) to $G$ from every problem in NP, or (equivalently) if there is a reduction from an NP-hard problem to $G$. A reduction from $F: I \to \{\text{yes}, \text{no}\}$ to $G: I' \to \{\text{yes}, \text{no}\}$, which we write $F \xrightarrow{\text{red}} G$, means that there is a polytime algorithm $f: I \to I'$ with

$$F(x) = \text{yes} \Leftrightarrow G(f(x)) = \text{yes}. \tag{1}$$

Any polytime solution algorithm for $G$, then, may be transformed into a polytime solution algorithm for $F$ via the reduction. If a polytime solution algorithm exists for $G \in$ NP-hard then it may be used to produce a polytime solution algorithm for any $F \in$ NP, in which case $P = NP$. If $P \neq NP$, then there is no polytime solution algorithm for $G$, and therefore any solution algorithm for $G$ must be exponential time or worse. A problem that is both in NP and NP-hard is called NP-complete.

A problem is in co-NP if its complement is in NP; co-NP hard and co-NP complete problems are defined analogously to the NP case. Colloquially, for a problem $F \in$ NP it is easy to verify that a proposed answer is indeed a solution; i.e., if $i \in I$ has $F(i) = \text{yes}$, then the proposition "$F(i) = \text{yes}$" may be verified in polytime. Conversely, for a problem $F \in$ co-NP it is easy to verify that a proposed nonanswer is indeed not a solution; i.e., if $i \in I$ has $F(i) = \text{no}$, then the proposition "$F(i) = \text{no}$" may be verified in polytime. It is widely believed [23], albeit unproven, that P, NP, and co-NP are mutually inequivalent classes, although trivially $P \subseteq NP$ and $P \subseteq$ co-NP. Note that if $P = $ co-NP then $P = NP$, and hence NP $=$ co-NP since P is closed under complement. Therefore, if $P \neq NP$

---

[1]Up to a Turing machine's theoretical infinite amount of memory, most programming languages are Turing complete, i.e., as powerful as a Turing machine.

then P ≠ co-NP, and there is no polytime algorithm for any co-NP-hard problem.

While it is *a priori* not clear that NP-complete problems exist, Cook [24] and Levin [25] discovered the first NP-complete problem SAT, which we now describe since it will be used later. A *literal* of a boolean variable $x$ is $x$ or its *negation* $\neg x$ (read "not x"), a *clause* is an *or* of literals (e.g., $x_1 \vee x_3 \vee \neg x_{10}$), and a *CNF-formula* is an *and* of clauses, (e.g., $x_1 \wedge (x_2 \vee \neg x_1) \wedge (\neg x_1 \vee \neg x_3)$). A CNF-formula $\varphi$ is *satisfiable* iff there is an assignment of values to the boolean variables such that $\varphi$ evaluates to yes. SAT is the problem of determining whether a given CNF-formula is satisfiable, and the Cook-Levin theorem shows that there exists a reduction $F \xrightarrow{\text{red}} \text{SAT}$ for all $F \in \text{NP}$. Since SAT is clearly NP, it is also NP-complete. Furthermore, SAT $\xrightarrow{\text{red}}$ CLIQUE, showing that CLIQUE is NP-hard. CLIQUE is also NP, and therefore NP-complete as well.

Often, complexity classes are discussed with regard to decision problems. While this may seem like a limitation, more general problems (such as optimization problems) can often be reformulated as decision problems. If the optimization problem is to minimize an objective function $h(x)$, the associated decision problem is to decide whether or not there is a point $x^*$ in the domain such that $h(x) \leq \zeta$, where $\zeta$ is additional data required to specify an instance of the decision problem. Alternatively, one may formulate a decision problem that decides whether a given point $x^*$ minimizes $h(x)$; this is the type of decision problem we predominantly study in this work. Intuitively, the optimization problem $O$ is at least as hard as its associated decision problem $D$: an algorithm that solves $O$ implicitly solves $D$. In some cases, an algorithm that solves $D$ may be turned into an algorithm that solves $O$.

## III. COMPLEXITY OF VACUA AND NEAR-VACUA

We now study the complexity of vacua and points in field space that are close to vacua. The input to the problems will be a scalar potential $V(\phi)$ for which all important contributions have been computed. Throughout, we use minimum and vacuum interchangeably. See Sec. IV for the complexity of determining these contributions.

We also note that, though the fields and parameters that appear in the scalar potential are in general complex, the real and imaginary parts combine to form a real function. Therefore many results from computer science and optimization, which often utilize real functions, can be directly applied to scalar potentials.

We proceed by reviewing results on the minimization of nonlinear functions, then turn to versions relevant for metastable and stable vacua, and finish with near-vacua.

*Minimization of scalar potentials* While (continuous) linear problems can be minimized in polynomial time, the computational complexity and decidability of nonlinear problems are much harder. A fairly general formulation is unconstrained NONLINEAR PROGRAMMING (NLP):

*NLP: Find the global minimum of $h(x)$,*

where $h(x)$ is a differentiable function in several unknowns. The associated decision problem is to decide whether a given point $x^*$ is a global minimum. These problems may also be modified by imposing constraints, such as equalities, inequalities or within intervals. In constrained problems it can happen that critical points are outside the feasible region (i.e., do not satisfy the constraints), in which case the function is minimized at the boundary of the feasible region and not at a critical point. However, these cases are usually not of physical interest due to being at the boundary of validity of the effective field theory (EFT), as we will discuss.

The hardness of NLP can be related to the hard problems that arise in the study of critical points, i.e., finding critical points and deciding whether the critical point is a (global) minimum. Both are relevant for the optimization version of NLP stated above, but only the latter is relevant for the decision version of the problem, since the former can be easily decided by evaluating $\nabla h(x^*)$, assuming derivatives can be computed.

First, consider the problem of finding a critical point. This problem requires solving the system of equations $\nabla h = 0$. If $h$ is a polynomial, finding a critical point requires finding a nontrivial root of a system of polynomial equations. We therefore first discuss the complexity of solving such systems: given a field $\mathbb{K}$ of arbitrary characteristic, it is NP-hard to decide whether a system of $s$ homogeneous polynomials in $\mathbb{K}[x_0, \ldots, x_n]$ has a nontrivial root [26]. Finding roots of nonhomogeneous polynomials is also NP-hard, as may be proven via reduction from SAT. Given an instance of SAT, which is a CNF-formula, form an associated system $S$ of non-homogeneous polynomials as follows. For each boolean variable $x_i$ of SAT, add $x_i(1 - x_i)$ to $S$. Associate a polynomial $p(l)$ to each literal $l$ via $p(l = \neg x) = x$ and $p(l = x) = (1 - x)$. To a clause $C$ that may appear in SAT, $C = l_1 \vee \ldots \vee l_n$, associate $\tilde{p}(C) = \prod_{l \in C} p(l)$; for example, $\tilde{p}(x_1 \vee x_2 \vee \neg x_3) = (1 - x_1)(1 - x_2)x_3$. For each clause $C$ in the CNF-formula, add $\tilde{p}(C)$ to $S$. The system of polynomials $S$ has a nontrivial root iff the CNF-formula is satisfiable. Thus, finding a nontrivial root of a system of polynomials is NP-hard.

Algebraically, simultaneous zeros of a system of equations can be found by computing resultants or Gröbner bases, see e.g., [27] for an application of the latter to finding string vacua. The above results show that computing resultants is NP-hard [28]; computing Gröbner bases is trivially hard because in the worst case there are known to be a doubly exponential number of Gröbner basis elements [29].

Finding a critical point of a function $h$ requires solving a system of equations related to $h$ by taking derivatives. From a physical point of view, the vacuum manifold of the scalars in the EFT is smooth in regions where the EFT can be trusted. Thus, the tangent space to points in such regions is full-dimensional and the system of polynomials obtained from the gradient is as hard to solve as any generic set of polynomials. However, we can also directly show NP-hardness of finding critical points by further reducing the problem of finding solutions of a system of polynomial equations to the problem of finding solutions to $\nabla h = 0$: Given a set of polynomials $\{f_i(x) = 0\}$, form a polynomial with additional variables $y_i$, $h(x,y) = y_i f_i(x)^2$. Then $h(x,y)$ has a critical point iff the system $\{f_i(x) = 0\}$ has a root. This proves that finding critical points of $h(x)$ is NP-hard.

Second, consider the problem of whether a given critical point $x^*$ is a (local or global) minimum. A critical points is a minimum if the Hessian is positive definite (PD), a maximum if the Hessian is negative definite, and a so-called *strict* saddle point if the Hessian has positive and negative eigenvalues. Ambiguity arises if the Hessian is positive semidefinite (PSD) but not PD, i.e., it has zero and a positive eigenvalues: a PSD Hessian is necessary for a minimum, but the critical point could also be a saddle point. Similarly, a negative semidefinite Hessian is necessary for a maximum, but again the critical point could be a saddle point. Since at these points some of the first two derivatives vanish, higher-order derivatives are needed to decide the nature of the critical point.

However, while saddle points with PSD Hessian can be abundant in polynomials with symmetries, they are of measure zero in general polynomials, such that almost all local minima have positive definite Hessians [30]; this is consistent with [31], since the many saddles in that work do not have PSD Hessian. The proof regarding the measure zero set uses elements from algebraic geometry (Zariski topology, elimination theory) and does not apply beyond polynomial functions [32,33]. Thus, the measure zero result does not apply to general NLPs.

A key result of Murty-Kabadi [34] is that finding even a *local* minimum of an unconstrained NLP is co-NP-hard; this is proven via the decision version of the problem. Even in the case that $h(x)$ is a quadratic polynomial, finding local minima is co-NP-hard [22,34,35]; we will discuss this and a caveat at length.

This hardness of the decision problem of local minima is perhaps surprising, so before performing a detailed analysis let us naively ask: what could possibly be hard about determining whether or not a given point $x^*$ is a local minimum for a general function $h(x)$? Since it is polytime to determine whether $x^*$ is a critical point and, if so, whether the Hessian is PD, the hard case for a critical point is necessarily when the Hessian there is PSD but not PD.

This is a physically interesting case. Letting the objective function $h(x)$ be $V(\phi)$ in an EFT, it requires the existence of a zero eigenvalue for the Hessian, which means that $V(\phi)|_{\phi^*}$ has a flat direction to quadratic order. Suppose there is only one. Then for $\phi^*$ to be a local minimum, either the flat direction remains flat to all orders and $\phi^*$ belongs to a moduli space, or the flat direction is stabilized by higher order terms. If the leading higher order term is negative, $\phi^*$ is a saddle point. Of course, there may be multiple zero eigenvalues, and the higher order terms determine the nature of the critical point at $\phi^*$. The conclusion for isolated vacua, which are of interest in the landscape, is that determining whether a critical point $\phi^*$ is a vacuum is difficult only if at least one of the directions away from $\phi^*$ is stabilized by terms of degree $> 2$. Such terms could be, e.g., D-terms in supersymmetric theories. However, since finding critical points (as in the optimization problem that nature solves) is already hard, deciding the nature of a critical point is only the second hard step in an already hard problem.

Let us close this subsection with a remark on the available optimization algorithms. With the advent of (supervised) machine learning, decisive progress has been made both in theoretical and practical investigations for minimizing NLPs. The most commonly used techniques are gradient descent (GD) techniques, where a minimum is found by following the direction of steepest descent, reminiscent of how energy is minimized in classical systems. There are several ways in which GD methods are improved and used in supervised machine learning, e.g., momentum GD, stochastic GD or noisy GD (see e.g., [36] for an overview).

Gradient descent (GD) minimization runs the risk of getting stuck at critical points with vanishing gradient that are saddle points rather than local minima. In such cases, momenta or noise can move you away from the zero gradient locus such that the GD can continue. The same mechanism (e.g., due to thermal or quantum fluctuations) prevents physical systems from getting stuck at local maxima or saddle points. Indeed, it has been shown that for functions with only strict saddle points, noisy or stochastic GD can find local minima in polynomial time [37]. There are also techniques to overcome saddle points at which (at most) the first three derivatives vanish [33], but the problem of finding fourth order local optima is NP-hard.

### A. Metastable vacua

Having introduced a very general version of a minimization problem, an unconstrained NLP, we now turn to the closely related problem of determining metastable vacua in an effective field theory. The problem is METASTABLE-VACUUM (MSVAC):

*MSVAC: Find a local minimum of $V(\phi)$, such that $l_i \leq \phi_i \leq u_i$,*

where $\phi = (\phi_1, \ldots, \phi_n) \in \mathbb{R}^n$ and $l_i, u_i \in \mathbb{R}$ are used to give so-called box constraints. Here the box constraints encode the fact that the EFT has a regime of validity that bounds the scalar fields values.

In order to prove that MSVAC is co-NP-hard, we first modify a proof of co-NP-hardness for the local quadratic programming problem QPLOC [22]. The proof utilizes a reduction from $\overline{\text{MAX} - \text{CLIQUE}}$ and is reproduced in Appendix. We take the original result and impose box constraints by replacing the general constraints $Ax \geq b$ with $l_i \leq x_i \leq u_i$. This may be done by taking the final instance of QPLOC in the Appendix proof and putting it on $\{0 \leq x_i \leq u_i\}$ with $u_i > 0$; the proof version was on $\{0 \leq x_i\}$, but the upper bounds just imposed do not change the validity of the relationship between cliques and whether or not $y^* = 0$ is a local minimum. The co-NP-hard decision problem we have just constructed is a subset of QPLOC with box constraints and lower bound 0:

*0-BOX-QPLOC: Is $y^* = 0$ a minimum of the function $f(y) = y^T H y$ for $0 \leq y_i \leq u_i$?*

Using this, we can construct a co-NP-hard decision problem for quartic programming with box constraints, utilizing an idea from Sec. IV of [34]:

*BOX-QUARTLOC: Is $x^* = 0$ a minimum of the function $g(x) = x_i^2 H_{ij} x_j^2$ for $-u_i \leq x_i \leq u_i$?*

In fact, via the transformation $y_i = x_i^2$, the point $x^* = 0$ is a local minimum of $g(x)$ iff $y^*$ is a local minimum of $f(y)$, which occurs iff $G$ does not have a clique of size $\geq k$. Following the logic through these various problems, this establishes that BOX-QUARTLOC is co-NP-hard by reduction from $\overline{\text{MAX} - \text{CLIQUE}}$. Therefore (the decision version of) MSVAC is co-NP-hard by reducing BOX-QUARTLOC to MSVAC.

We would like to make a number of comments about the choices we have made in constructing BOX-QUARTLOC and its relation to results in the optimization literature.

The astute reader may ask why we mapped the problem to a quartic instance of MSVAC, as opposed to a quadratic instance via the inclusion map. First there are simplifications of the decision problem for isolated minima of quadratic functions, as discussed below, but also there is the complication that the point $\phi^* = 0$ would be on the boundary of the domain, and in field theory we are interested in vacua that are away from the boundary to ensure that we are within the regime of validity of the EFT. In fact the hardness of QPLOC is derived entirely from such boundary points: all three proofs [22,34,35] that QPLOC is co-NP-hard utilize such boundary points, and determining whether or not an interior point $x^*$ is a local minimum of $x^T Q x + c^T x$ is in P. For this reason we have reduced to BOX-QUARTLOC, for which determining whether or not an interior point is a local minimum is co-NP-hard.

More generally, one can inquire about instances of MSVAC with degree $n$ polynomial $V(\phi)$ and $\phi^*$ an interior point. There, even deciding whether a polynomial is convex, which is useful for minimization, is NP-hard for even degree larger than 2; it is P for odd degree [38].

In summary, co-NP-hard instances of the decision version of MSVAC have a PSD Hessian at $\phi^*$ that is not PD. Such a point has flat directions in the potential to quadratic order; higher order terms may stabilize or destabilize the vacuum. In addition, the optimization version of MSVAC also requires *finding* a critical point, which itself is an NP-hard problem.

### B. Stable vacua

Let us now determine the complexity of finding stable vacua. These are vacua for which down-tunneling is not possible, and in this sense they are more stable than local minima, for which down-tunneling is often possible. We will call such vacua stable vacua, keeping in mind that this is a bit of a misnomer due to the possibility of up-tunneling.

We define STABLE-VACUUM (SVAC):

*SVAC: Find a global minimum of $V(\phi)$, such that $l_i \leq \phi_i \leq u_i$,*

where $\phi = (\phi_1, \ldots, \phi_n) \in \mathbb{R}^n$ and $l_i, u_i \in \mathbb{R}$. Since finding global minima is at least as hard as finding local minima, SVAC requires first solving an NP-hard problem to find critical points, subsequently solving a co-NP-hard problem to determine whether these points are minima, and then selecting the global minimum.

### C. Complexity of near-vacua and rolling

We have studied the complexity of finding stable and metastable vacua. We have shown that both finding critical points and deciding whether a critical point is a minimum is computationally hard via reduction from SAT and $\overline{\text{MAX} - \text{CLIQUE}}$, respectively. In this section, we turn to study the complexity of finding points in field space that are approximate local minima, i.e., where gradients of the potential are small but not necessarily zero. Concretely, we say that $x^*$ is an $\epsilon$-*approximate local minimum* of a continuous function $f: U \to \mathbb{R}$ if there exists an open set $N$ containing $x^*$ such that $f(x^*) \leq f(x) + \epsilon |x - x^*|$ for all $x \in N \cap U$; here $dim(U) = n$. That is, $x^*$ may not be a local minimum, but it is close to one. Henceforth, such points are called $\epsilon$-minima or $\epsilon$-vacua for brevity. We thus have the problem:

*NEAR-VAC: Given a scalar potential $V(\phi)$ and $\epsilon > 0$, find a point $\phi^*$ that is an $\epsilon$-vacuum.*

While general NLPs have no polynomial time algorithm if P ≠ NP, $\epsilon$-minimizing arbitrary differentiable functions is *linear* in $n$ under mild assumptions. Before stating the theorem, we need the notion of a Lipschitz condition. The function $f$ is said to obey a *Lipschitz condition with bound $M$* if $|f(x) - f(y)| \leq M|x - y|$ for all $x, y$ in $U$. Then [39]:

**Theorem:** Let $f: U \to \mathbb{R}$ be a $C^1$ function whose gradient satisfies a Lipschitz condition with bound $M$. Then an $\epsilon$-minimum can be found with at most $4n(M/\epsilon)^2$ function and gradient evaluations.

The upper bound on the running time of the algorithm diverges as $\epsilon \to 0$, i.e., in the limit that $\epsilon$-minima are definitively local minima. This is consistent with co-NP-hardness of local minima, since otherwise a solution algorithm for local minima that is linear in $n$ would exist. See [39] for an explicit algorithm for finding $\epsilon$-minima with this runtime.

For practical purposes, note that any computer only uses finite precision $\delta$, and therefore any vacuum minimization problem that it solves effectively solves NEAR-VAC with $\epsilon \geq \delta$. Smaller $\epsilon/M$ (which is the step size in the algorithm) requires higher precision, and therefore since finding the exact minimum using this algorithm requires taking $\epsilon \to 0$, it also requires infinite precision. However, even at infinite precision, the computation time of this algorithm diverges and the problem is not solved.

By applying this result to a scalar potential $V(\phi)$, we conclude that there exists a classical algorithm for finding $\epsilon$-minima of $V(\phi)$ that is linear in the number of scalar fields $n$. Such a solution is in general a rolling solution. The implications of this result will be discussed below.

## IV. COMPLEXITY OF STRING POTENTIALS

To even begin the computationally hard minimization problems above, the important contributions to $V(\phi)$ must be determined. We now study the complexity of determining some such contributions in string theory.

We will mostly organize our discussion based on known computationally hard problems and demonstrate that problems in string theory related to scalar potentials are instances of these problems. This is potentially weaker than showing that the string theoretic problems themselves are computationally hard, since it is possible in principle that the set of instances realized in string theory make up a subset of the hard problems that is itself in P. However, we have no reason to think this is the case in the string theoretic problems below. Except for the first problem, which is NP-hard, the other problems below are all known to be NP-complete; see [40] for further discussion.

In order to write down the scalar potential as derived from string theory, we need to know the light (relative to the Kaluza-Klein scale) scalar fields:

*STRING-SPECTRUM: Given a string background, find all light scalar fields.*

While the way of computing these depends on the string theory in question, the problem often requires solving hard problems. In F-Theory, computing the singlet spectrum often requires resultants and/or Gröbner basis computations in order to ensure that ideals corresponding to matter loci are not overcounted [41]. As discussed in Sec. III, these computations are NP-hard. Moreover, neutral singlets at singularities can be counted from the Milnor or Tjurina number of the singularity [42], whose computation requires reducing equations modulo ideals and thus usually also require Gröbner basis computations. In heterotic theories on smooth Calabi-Yau manifolds with vector bundles, computing the spectrum requires evaluating bundle cohomologies. This is (even for line bundles) exponential in the number of generators of the Stanley-Reisner ideal [43], which grows with the number of geometric moduli. However, see [44] for speed-ups in specific setups, [6] for the first application of machine learning to speed up line bundle cohomology computations, and [45] for another recent machine learning application to this subject.

The next three problems we review are known to be computationally hard. We demonstrate how instances of them arise in scalar potential calculations.

Given a graph $G$, lengths $l(e) \in \mathbb{Z}^+$ for all edges $e \in E$, a subset $E' \subset E$, and a bound $B \in \mathbb{Z}^+$, we have the problem:

*RURAL POSTMAN: Is there a circuit (closed loop) in $G$ that includes each edge in $E'$ and that has total length no more than $B$?*

This problem is NP-complete in general, and also if edge lengths are set to one or if the graph is directed.

Consider a quiver gauge theory represented by $G$, and assume (as in supersymmetric theories) that there is a charged scalar field associated to each edge. Each circuit corresponds to a gauge invariant operator (GIO) $\mathcal{O}$ that may appear in the scalar potential. A physically relevant question is whether there exists a GIO that couples a fixed subset $E'$ of fields to one another, perhaps involving additional fields beyond those in $E'$. For this operator to be important, it is natural to ask for an upper bound $dim(\mathcal{O}) \leq B$. Putting this in graph theoretic language, it is precisely an instance of RURAL POSTMAN: such a GIO exists if and only if there is a circuit of maximal length $B$ that contains each $e \in E'$.

Now we turn to INTEGER PROGRAMMING (INT-PROG). Given $(A \in \mathbb{Z}^{m \times n}, b \in \mathbb{Z}^n, c \in \mathbb{Z}^n, B \in \mathbb{Z})$,

*INT-PROG: Is there a $y \in \mathbb{Z}^n$ such that $Ay \leq b$ and $c \cdot y \leq B$?*

This is asking whether or not there is an integral point that satisfies systems of hyperplane constraints, where the latter may cut out cones or polyhedra.

This problem is relevant for instanton corrections through the relationship between their zero modes and (in many cases) line bundle cohomology on toric varieties. The latter (see, e.g., [46]) requires counting the number of lattice points that satisfy hyperplane constraints (including for local sections, not just global sections), which requires determining whether such points even exist; i.e., it is an instance of INT-PROG. Computing line bundle cohomology on toric varieties also arises in many other contexts in string theory, for example in computing matter spectra of low energy gauge theories.

Finally, consider the problem QUADRATIC DIOPHANTINE EQUATION (QDE):

> QDE: Given a quadratic diophantine equation, does it have a solution?

Though generic diophantine equations are undecidable [47] (see [11] for a study of diophantine undecidability in string theory), any single QDE is decidable [48]. However, already in the case where the QDE has the form $ax_1^2 + bx_2 = c$, it is NP-complete [49].

Diophantine equations are ubiquitous in string theory, and QDEs also arise. Let us study Euclidean D3-instantons (ED3) on divisors $D$ on a Calabi-Yau threefold $X$, due to their relevance for the scalar potential. Consider potential zero modes that arise at the curve $C = D \cdot D_{D7}$ where the ED3 intersects one or more D7-branes. This zero mode sector may change the structure of the superpotential correction or cause it to not contribute at all. The study of the instanton correction is simplified in the absence of these zero modes, e.g., when $C = \mathbb{P}^1$ and there are no induced fluxes on $C$. Determining divisors $D$ such that $C$ is a $\mathbb{P}^1$ requires $\chi(C) = -D \cdot D_{D7} \cdot (D + D_{D7}) = 2$, which becomes a quadratic diophantine in the integer parameters of $D$ upon application of the triple intersection form. In [11] it was shown that these QDEs factorize into a solvable product of linear factors if $X$ is an elliptic fibration with single section; such cases are in P, but the general problem is NP-complete. ED3-instantons are hard for other reasons, as well, since resultants arise in the computation of Pfaffian prefactors [50].

With these examples, we have likely only scratched the surface of computationally hard problems that have instances arising in the computation of scalar potentials in string theory. For example, it would be interesting to study the complexity of flux superpotentials.

## V. DISCUSSION

We have shown that the problem of finding global and local minima of a scalar potential are both co-NP-hard. This means that, under the assumption that P ≠ NP, there is no polynomial time algorithm for finding vacua of a generic scalar potential. Furthermore, we demonstrated that even determining the scalar potential in string theory requires solving instances of known computationally hard problems.

The decision and optimization versions of the problems of finding metastable (or stable) vacua differ slightly in which aspects of the problem are hard. Both versions require determining whether a critical point $\phi^*$ is a local minimum, which may only be hard if the Hessian of $V(\phi)$ at $\phi^*$ is positive semi-definite (PSD); determining whether $\phi^*$ with positive definite (PD) Hessian is a local minimum is in P. In the PSD case the nature of the critical point is determined by terms beyond quadratic order. However, in the optimization version of the problem critical points must also be found, which itself is an NP-hard problem.

Of course, it is the optimization version of the problem that nature attempts to solve, and thus the added hardness of finding the critical point is important.

In the upcoming subsections of this discussion, we will consider only the implications of complexity arising from the hard problems discussed above, rather than from potential precision issues needed to compute vacua or their expectation values on a finite precision computer. Indeed, the above problems would be hard even on an infinite precision computer. For vacua in which the scalar field values cannot be represented with finite precision, minimization problems can become even more intractable. More specifically, for optimization problems, assuming finite precision calculations the only problems that can potentially be solved for general vacua is NEAR-VAC, unless the field values at the minimum may be represented on the finite precision computer. On the other hand, if the field value $\phi^*$ is not of this type, decision problem instances that utilize $\phi^*$ cannot even be formulated because it would require higher precision than is available on the computer. Therefore, some instances of the problems may become undecidable or ill-posed.

### A. Practical implications

Together, the nested hardness results of computing and subsequently minimizing scalar potentials explain a fact of string landscape studies: there are currently no concrete examples of metastable vacua at large numbers of moduli (e.g., $h^{11}$ or $h^{21}$ of at least $\mathcal{O}(10)$ in Calabi-Yau compactifications), despite the fact that the vast majority of metastable vacua are expected to be in such regions. The effective potentials there are either too difficult to construct or too difficult to minimize.

Let us comment on the de Sitter swampland conjecture in light of these results. Since finding minima of an objective function (a goal in the optimization community) does not depend on the value of the function at these points, results from the optimization literature do not address the question of whether a given vacuum occurs above or below zero, and thus cannot be directly applied to the dS swampland conjecture. However, the co-NP-hardness of MSVAC implies that it is also computationally hard to prove that a given point is *not* a local minimum, and therefore proving the absence of local minima, including positive ones, is also computationally hard. Both of these issues makes a direct verification of the conjecture even more intractable than the number of string compactifications in string theory already had; an indirect approach is necessary for verification. On the other hand, falsification requires only a single explicit counterexample. In light of the conjecture [15], simple counterexamples have been proposed, see, e.g., [17].

### B. Cosmological implications

We close with a discussion of the potential cosmological implications of the complexity of vacua and near-vacua.

Physically, an extremum of the scalar potential is reached (classically) by scalar fields $\phi_i$ moving along their trajectories as obtained from solving their classical equation of motions. The extremum may be a minimum, maximum, or saddle point, though the latter two cases require a fine-tuning of initial conditions that happens rarely; generic initial conditions lead to a minimum. The time it takes the fields to reach the extremum is thus a function of the potential and the initial conditions. This algorithm employed by a classical universe is reminiscent of minimization via momentum gradient descent.

However, there are of course thermal and quantum fluctuations, as well as quantum tunneling, which take the theory out of saddle points and maxima in a way reminiscent of *stochastic* gradient descent in supervised machine learning. This is not a classical algorithm (i.e., is not a deterministic Turing machine) and therefore lies outside the realm of classical complexity theory. A rigorous determination of the cosmological implications of this work therefore requires reevaluating its questions in the context of quantum complexity theory. Such a study is outside the scope of this work, but we emphasize that quantum algorithms do not necessarily result in a large speed-up; see, e.g., [10] for discussion and references.

Quantum dynamics in the universe can be thought of as a process or algorithm that optimizes the scalar potential, i.e., searches for minima, and if our classical hardness results persist into the quantum regime, it is interesting to consider their implications for cosmology in the context of a recent measure proposal [14]. There, the authors propose that we ended up in the universe we observe not necessarily because it is ubiquitous, but because it is easy to find, in the sense of computational complexity. This connects complexity classes of string vacua and problems that arise in them with a notion of vacuum likelihood. We call this the complexity measure.

We would like to consider our results in light of the complexity measure, under the assumption that our classical complexity analysis holds in the quantum regime.

Since SVAC and MSVAC are co-NP-hard, and therefore in general finding vacua is exponentially hard, the algorithm used by the universe to end up in a vacuum (a version of stochastic gradient descent with tunneling) cannot perform the task in polytime unless P = NP. It is then tempting to conclude from the complexity measure that our universe (or bubble in a multiverse) is in a rolling solution. This may, in fact, be the case, but there are two important caveats that we would like to discuss.

The first caveat is that this conclusion has not taken into account vacuum lifetimes. To make this more precise, suppose the universe or a patch in the multiverse is born with some initial conditions that lead to rolling on a scalar potential. Eventually it will stop at a vacuum in a stopping time $t_s$, where the vacuum has lifetime $\tau$. Our hardness result implies that there are vacua for which

$$t_s \sim e^N, \qquad (2)$$

where $N$ is the number of scalar fields. If $\tau$ scales as $\tau \sim e^{aN}$ with $a \gg 1$, then $\tau/t_s \gg 1$ and the universe (or bubble) spends most of its time in the vacuum, which is what we should expect to see. On the other hand, if $\tau$ is polynomial in $N$ or if $a \ll 1$ then the complexity measure prefers a rolling solution. Resolving this issue requires detailed knowledge of lifetimes (as a function of $N$) in large ensembles of vacua. Such a study is beyond the scope of this work, but is an interesting direction due to its implications in the complexity measure.

The second caveat is that the hardness result, which ensures the existence of vacua with $t_s \sim e^N$, does not preclude the existence of vacua in P. If such vacua exist, and if nature's minimization procedure is a polytime algorithm on those vacua, then these vacua are preferred in the complexity measure and there is no clear complexity reason to prefer solutions that are rolling towards a vacuum over solutions that have already settled in the vacuum. In field theory, models with P-vacua include those where $V(\phi)$ is a quadratic polynomial. However, such models are unlikely to occur in compactifications of string theory with $\mathcal{N}=0$ or $\mathcal{N}=1$ supersymmetry, since Planck-suppressed operators and exponentials generically arise from gravitational effects and instanton contributions, respectively. More generally, the existence of P-vacua seems to require $V(\phi)$ to have very special structure that (in our opinion) is unlikely to exist in $\mathcal{N}=1$ or $\mathcal{N}=0$ string compactifications, due to the spoiling of special structure by, e.g., instantons and gravitational effects.

The hardness of finding vacua given $V(\phi)$ is important in the complexity measure, but it is also important to determine whether there are string theory backgrounds in which determining $V(\phi)$ is easier than in other backgrounds. Though the general problem of determining $V(\phi)$ is conjectured to be undecidable [11], this does not preclude the existence of backgrounds in which determining $V(\phi)$ is in P. For example, as mentioned above, aspects of computing instanton contributions for elliptically fibered Calabi-Yaus is in P while the general problem is hard, making these Calabi-Yaus more likely in the measure of [14] (at least if we only consider the computation time to find the Euclidean instanton corrections to the superpotential). It would be interesting, to speculate what this means for Calabi-Yau manifolds that are elliptically fibered, but have mirrors that are not.

In conclusion, in addition to the potential implications for rolling solutions, our results strengthen the preference of the complexity measure for vacua at small $N$. This should be contrasted with the fact that most vacua are expected to exist at larger $N$, and a recent study [9] that demonstrates preferential vacuum selection at large $N$ in another measure [51]. Resolving this tension is an important direction for future work. One interesting possibility is if there is important physics that is easier to compute at large $N$, which balances the hardness of minimization at large $N$.

## ACKNOWLEDGMENTS

We thank Frederik Denef, Michael Douglas, Cody Long, Liam McAllister, and Scott Watson for discussions. J. H. thanks the Simons Center for Geometry and Physics for hospitality and participants of the Simons Summer Workshop for discussions. J. H. is supported by NSF Grant No. PHY-1620526. The work of F. R. is supported by the EPSRC network Grant No. EP/N007158/1.

## APPENDIX: PROOF THAT QPLOC IS CO-NP-HARD

In this Appendix we repeat the proof of [22] that finding local minima of a quadratic polynomial objective function with linear constraints is co-NP-hard. We will need a modification of this proof to establish that MSVAC is computationally hard. It also demonstrates a polytime reduction.

A linearly constrained NLP that has a quadratic polynomial $h(x)$ is known as quadratic programming (QP). Stated as an optimization problem, QP is the problem:

*QP: Find a global minimum of $x^T H x + c^T x$ subject to $Ax \geq b$,*

where $x \in \mathbb{R}^n$, $b \in \mathbb{R}^m$, $H$ is a symmetric $n \times n$ matrix, and $A$ is a $m \times n$ matrix. An instance of this problem is defined by $(H, c, A, b)$, and the output is the global minimum. The associated decision problem has an additional input $\zeta$, and asks whether there is an $x$ such that $Ax \geq b$ and $x^T H x + c^T x \leq \zeta$. The general QP problem is NP-complete.

QP has important special cases. In the case that $H$ is positive semi-definite the objective-function is convex and QP is easier to solve; this problem CONVEX-QP is in P. QP may also be endowed with box or simplex constraints, in which the general constraints $Ax \geq b$ are replaced by box constraints $l_i \leq x_i \leq u_i$ and simplex constraints $x \in \Delta^{n-1} = \{(x_1, \ldots, x_n) | \sum x_i = 1, x_i \geq 0\}$, respectively. These last two modifications, which we call BOX-QP and SIMPLEX-QP, will both be relevant later. Despite the fact that both problems are smaller than QP, they are still NP-complete. See the book [22] for the results in this paragraph and their original references.

The local version of a decision problem for QP is QPLOC:

*QPLOC: Given an instance $(H, c, A, b)$ of QP and a point $x^* \in \mathbb{R}^n$, does $x^*$ satisfy $Ax^* \geq b$ and, if so, does there exist an $\epsilon$ such that $\frac{1}{2} x_i A_{ij} x_j + c_i x_i \geq \frac{1}{2} x_i^* A_{ij} x_j^* + c_i x_i^*$ for all $x$ satisfying $|x - x^*| \leq \epsilon$ and $Ax \geq b$?*

That is, is $x^*$ a local minimum of this QP instance? QPLOC is co-NP-complete [34]; other proofs also exist [22,35].

We show that QPLOC is co-NP-complete by reduction of the co-NP-complete problem $\overline{\text{MAX} - \text{CLIQUE}}$, the complement of the NP-complete MAX-CLIQUE. Given a graph $G$ and $n \in \mathbb{Z}^+$, MAX-CLIQUE asks:

*MAX-CLIQUE: Does $G$ have a clique of size $\geq n$?*

The proof is as follows. Given an undirected graph $G$, $k \in \mathbb{Z}^+$, and $x_i \in \mathbb{R}$, consider the $G$-dependent quadratic

$$f(x) = -\sum_{(i,j) \in E(G)} x_i x_j =: x^T H_0 x, \quad (A1)$$

where $E(G)$ is the set of edges of $G$ and $(i, j) \in E(G)$ means that there is an edge between nodes $i$ and $j$. Consider also $g(x) = f(x) - c$ and $h(x) = f(x) - c(\sum x_i)^2 =: x^T H x$, where

$$c = \frac{1}{2k-1} - \frac{1}{2}. \quad (A2)$$

Note that $g(x)$ and $h(x)$ are identical functions on the simplex $\Delta^{n-1} = \{(x_1, \ldots, x_n) | \sum x_i = 1, x_i \geq 0\}$. By the Motzkin-Straus theorem [52], the optimum value of $f(x)$ on $\Delta^{n-1}$ is

$$\min(f(x)) = \frac{1}{2k} - \frac{1}{2}, \quad (A3)$$

where $k$ is the maximum clique size of $G$. Using this lemma and $h(x) = g(x)$ on $\Delta^{n-1}$, it follows that $h(x) = x^T H x$ has a negative value on $\Delta^{n-1}$ iff MAX-CLIQUE$(G, k)$ is yes, i.e., $G$ has a clique of size at least $k$. We have accomplished finding a quadratic objective function $h(x)$ depending on $(G, k)$ that is related to MAX-CLIQUE; i.e., we are moving towards a $(G, k)$-dependent instance of QPLOC, as necessary for reduction.

Now consider the related problem to minimize $y^T H y$ for $y \in \mathbb{R}^n$ subject to $y \in Q := \{y \in \mathbb{R}^n | y_i \geq 0\}$ and $\sum y_i = a > 0$. The input for this problem is $(G, k, a)$, and by scaling the problem as $x_i = y_i/a$ it may be related to the problem on the simplex, so $h(y) = y^T H y$ has a negative value on $Q \cap \{\sum y_i = a\}$ iff MAX-CLIQUE$(G, k)$ is yes; for fixed $(G, k)$ the answer to the decision problem is the same for any value of $a$.

Finally, we arrive at a $(G, k)$-dependent instance of QPLOC, as required for the reduction: is $y^* = 0$ a local minimum of $h(y) = y^T H y$ for $y \in Q$? Let us investigate this question with respect to cliques. Suppose $G$ does not have a clique of size $\geq k$; then $y^T H y$ is non-negative on $Q$ by the preceding argument and therefore 0 is a local lminimum. On the other hand, suppose $G$ has a clique of size $\geq k$; then there exists a $y_n$ such that $y_n^T H y_n$ is negative, and via the scaling $y_n$ can be brought arbitrarily close to $y^* = 0$, in which case it is not a local minimum. These two facts show $y^*$ is a local minimum iff $G$ does *not* have a clique of size $\geq k$. The latter is an arbitrary instance $(G, k)$ of $\overline{\text{MAX} - \text{CLIQUE}}$, and the former is an associated instance of QPLOC. This polytime reduction proves that QPLOC is co-NP-hard [22].